\documentclass{article}
\usepackage{arxiv}

\usepackage[utf8]{inputenc} 
\usepackage[T1]{fontenc}    
\usepackage{hyperref}       
\usepackage{url}            
\usepackage{booktabs}       
\usepackage{amsfonts}       
\usepackage{nicefrac}       
\usepackage{microtype}      
\usepackage{lipsum}		
\usepackage{graphicx}
\usepackage{natbib}
\usepackage{doi}

\usepackage{amssymb}
\usepackage{epstopdf}
\usepackage{hyperref}
\usepackage{listings}
\usepackage{color}
\usepackage{lscape}

\definecolor{javared}{rgb}{0.6,0,0} 
\definecolor{javagreen}{rgb}{0.25,0.5,0.35} 
\definecolor{javapurple}{rgb}{0.5,0,0.35} 
\definecolor{javadocblue}{rgb}{0.25,0.35,0.75} 
 
\title{Probability of Insect Capture in a Trap Network: Low Prevalence and Detection Trapping with TrapGrid}

\author{ \href{https://orcid.org/0000-0001-5062-7256}{\includegraphics[scale=0.06]{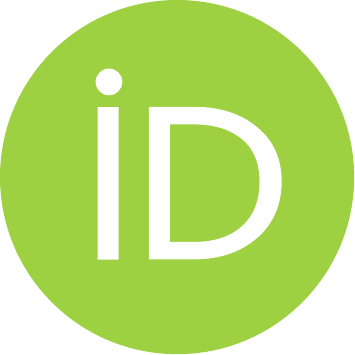}\hspace{1mm}Nicholas C. Manoukis}\thanks{Corresponding author. 64 Nowelo St. Hilo Hawai'i USA 96720} \\
	Tropical Crop and Commodity Protection Research Unit\\
	Daniel K. Inouye U.S. Pacific Basin Agricultural Research Center\\
	Hilo Hawai'i USA \\
	\texttt{nicholas.manoukis@usda.gov} \\
	\And
	\href{https://orcid.org/0000-0002-7762-5258}{\includegraphics[scale=0.06]{orcid.pdf}\hspace{1mm}Matthew P. Hill} \\
	Data61\\
	CSIRO\\
	ACT, Australia\\
	\texttt{matt.hill@csiro.au} \\
}

\renewcommand{\shorttitle}{\textit{TrapGrid} Low prevalence, detection}

\hypersetup{
pdftitle={A template for the arxiv style},
pdfsubject={q-bio.NC, q-bio.QM},
pdfauthor={Nicholas C. Manoukis et al},
pdfkeywords={Insect, Invasive pests, lures, Surveillance},
}

\begin{document}
\maketitle

\begin{abstract}
	Attractant-based trap networks targeting insects are ubiquitous worldwide. These networks have diverse targets, goals, and efficiencies, but all are constrained by practical considerations like cost and available lures. An important way to balance goals and constrains is through quantitative mathematical modeling. Here we describe an extension of a computer model of trapping networks known as ``TrapGrid'' to include an alternative mode of calculating the probability of capture over time in a trapping network: Strict detection (``capture of one or more'') compared with the average probability of capture as implemented in the original version. We suggest that this new calculation may be useful in situations of low prevalence where trap network operators wish to interpret the meaning of zero captures at a small scale. The original remains preferred for comparing the sensitivity and suitability of alternate trap networks (i.e. density of traps, their placement, lure attractiveness, etc). 
\end{abstract}


\section{Introduction}

Invasive insect pests are a critical and growing threat to agriculture, non-human dominated ecosystems, economic sustainability, and human health worldwide \citep{bradshaw_massive_2016,hill2016drivers}.  
There is a growing appreciation that management and mitigation of the risks posed by invasive species, including insects, is a key consideration for the continued development of international trade \citep{epanchin-niell_biological_2021}. For some of the most serious insect pests we can count on chemical lures that can be highly effective at bringing the target pest to a trap, where it is able to be captured. 

When the pest is of significant economic, environmental, or other significance and the lure sufficiently effective, trap networks are regularly created and maintained, often by government entities. These may be designed to detect an invasive pest that is not established in the area (``detection'' or ``surveillance'') or may be used to monitor a standing population of an insect (``prevalence''). An instructive example is the tephritid fruit fly \emph{Ceratitis capitata}, informally known as the Mediterranean fruit fly (``Medfly''), a global pest of high significance for hundreds of horticultural crops. In California U.S.A, there is a network of over 30,000 traps baited with a male lure dedicated to the detection of Medfly alone, and this has been an essential tool in keeping this species from becoming established and widespread in the state despite repeated incursions since 1974 \citep{gilbert_insect_2013, carey_year_2017, mcinnis_can_2017}. In Western Australia, where Medfly has been established for over a century, a permanent surveillance network is in place to monitor this pest, whilst in South Australia a trapping network is required to detect intermittent incursions \citep{broughton2017evaluation}. These traps form part of the 25,000 that comprise the national trapping grid for exotic and endemic fruit flies in Australia. 


Besides the expense and logistical issues that are inherent to operating them, it is not intuitively clear how effective a given trap network might be. Decisions like how many traps to deploy per unit area, where to site them, how often to check and service each trap can be informed by biological research, but remain difficult problems for all programs. One of the most effective way to address such issues has been through the use of mathematical modeling; and there is an extensive history of modeling to quantify and specify trap networks against agricultural pests (e.g. \cite{miller2015trapping}). 

A recent model to determine trap network sensitivity has been implemented in a computer program known as TrapGrid \citep{Manoukis2014Computer, HallTrapGrid}. While that program was originally intended to focus on surveillance scenarios, there is no reason the basic modelling approach could not be extended to other trapping situations such as population size estimates, with minor modification. Application to other situations might require changes to calculations, and also likely implies some changes in assumptions, or new assumptions. 

In this paper we describe an extension of the TrapGrid model to better analyze trap captures in scenarios of low prevalence compared with determining the average sensitivity of a trap network for purpose of optimizing surveillance. The latter was the original goal of TrapGrid, but the former applies to trapping scenarios worldwide. In addition to describing the new calculation, we execute a number of simulations to compare results side-by-side and offer suggestions on when one or the other approach might be desirable. A version of Trapgrid with both modes implemented and selectable via command-line switch is available for download at  \url{https://github.com/manoukis/TrapGrid}

\section{Methods}

\subsection{Modeling Overview}

Broadly speaking, TrapGrid originally aimed to answer two questions: 1) what is the average detection ability of a given trapping network and 2) what is the average probability of an insect from an invading population (that is not changing in size) being caught in a trap over time? As mentioned above, the original formulation was aimed at detection scenarios, where the average is an appropriate measure of the suitability and efficiency of the trap network. However, in situations where the target insect is established and at low prevalence, a different measure might be desirable; specifically 3) What is the probability of capturing one or more insects given the trap network and 4) how does the probability of capturing one or more insect increase over time? 

A full description of the model is in \cite{Manoukis2014Computer}, but a quick sketch is given here. For an individual trap the hyperbolic secant is used to relate distance from a trap to probability of capture (``capture probability model''). This can be calculated for several traps given positions of the traps and a given point in the arena as 1- probability of escape from all traps. This calculation can be run for a number of randomly chosen individual positions in the arena for an estimate of ``instantaneous'' capture probability. However, in order to accommodate a more realistic representation of a single population in space over time, TrapGrid allows modeling ``probability of capture over time''. This uses diffusion or random correlated walk models to simulate the spatial distribution of insects over time given an outbreak location and movement parameters. It is important to stress that under diffusion insect positions are re-plotted each time step (a day) and there is no continuous movement of individual particles. 

Here we are concerned especially with how cumulative probabilities are calculated- details are below for each mode. To compare results we re-create the main points of Fig. 6 in \cite{Manoukis2014Computer}. This figure shows the average capture probability at the end of two weeks of trapping and how it responds to varying trap densities and individual trap attractiveness. In this document we call this ``TGO''. The modified version (``TrapGrid-Alternative'') we refer to as ``TGA''. 



\subsubsection{Original TrapGrid Algorithm}

For each time step (``day'') of the simulation, the TGO algorithm calculates the \emph{average escape probability} for a set of insect positions. Positions are (usually) calculated via diffusion based on time since start and $D$, the diffusion coefficient. Once the average for the current day is calculated, the cumulative escape probability to the current day is calculated via the product of the previous day's cumulative value and the current day's average escape probability. In pseudo-code:

\begin{verbatim}
for each simulation;
 for each day;
  escapeProbability = 0;
  positions = distributeInsectPositions;
  for each position;
   escapeProbability = escapeProbability + escapeProbability(position);
  end for;
  dailyEscapeProbability = escapeProbability/numberInsectPositions;
  cumulativeEscapeProbability = cumulativeEscapeProbability * dailyEscapeProbability;
 end for;
 addResultsToOtherSimulations;
end for;
\end{verbatim}

\noindent Note that the initial escape probability for the day is 0, to accommodate averaging. This might lead to edge cases (where a single insect is used, for example, and it is randomly placed in the same position as a trap so its escape probability is also 0) where the program might result in a division error, although this is highly unlikely.

\subsubsection{Alternative Algorithm}

For each day of the simulation, the TGA algorithm calculates the \emph{probability all insects escape}, using positions calculated the same way as TGO, via the product of the individual insect escape probabilities. Cumulative probability over days is calculated as above. Pseudo-code:

\begin{verbatim}
for each simulation;
 for each day;
  escapeProbability = 1;
  positions = distributeInsectPositions;
  for each position;
   escapeProbability = escapeProbability * escapeProbability(position);
  end for;
  cumulativeEscapeProbability = cumulativeEscapeProbability * escapeProbability;
 end for;
 addResultsToOtherSimulations;
end for;
\end{verbatim}

\noindent Here the initial assumption is all flies escape, and that probability is modified by the addition of insect positions relative to traps. This is intuitively appealing if we are interested in the probability of capturing one or more insects. 

\section{Results}

\subsection{TGO and TGA Comparison}

\begin{landscape}
\begin{figure}[h]
\begin{center}
\includegraphics[width=8.7in]{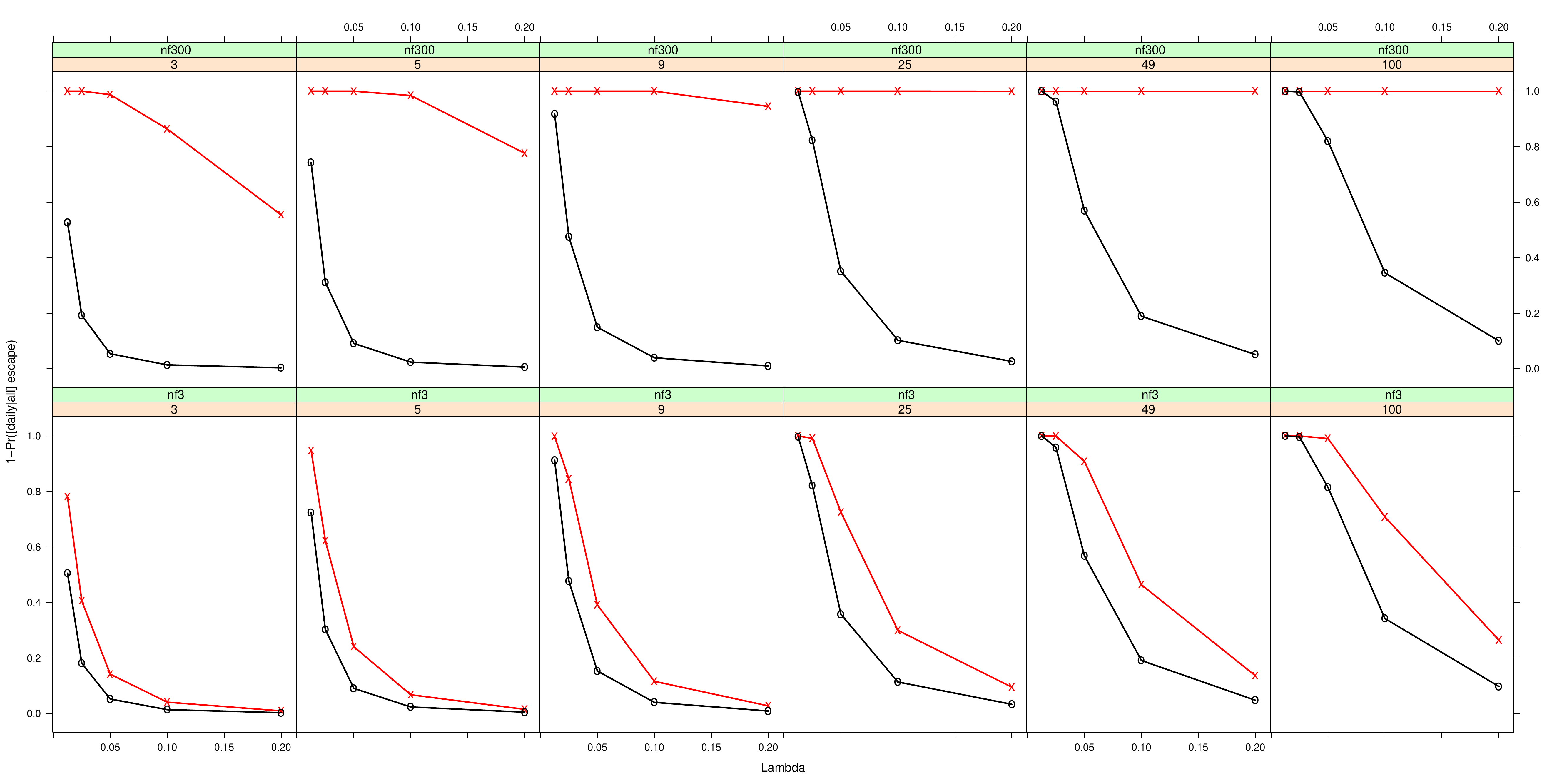}
\caption{Average capture probability (ordinate axes) as a function of trap attraction ($\lambda$, abscissa), number of traps (top row panel labels), number of insects (bottom row panel labels, nf300 = 300 insects; nf3 = 3 insects), and algorithm (black ``o'' = TGO; red ``x'' = TGA). Capture probability is ``daily average'' for TGO and ``of one or more'' for TGA.}
\label{fig:means}
\end{center}
\end{figure}

\end{landscape}

Systematic comparison employing each of the two algorithms, 6 levels of trap density (3, 5, 9, 25, 49, and 100 traps per square mile), 5 levels of trap attraction ($\lambda$ between 0.200 and 0.0125) and two insect population sizes ($N =$ 3 or 300) with 250 repetitions each results in a total of 30,000 simulations. Averages across the 250 simulations are shown in Fig. \ref{fig:means}.

Differences between TGO and TGA are clear when the number of insects is 300; at 3 insects, results are largely comparable: there is a similar impact of $\lambda$ (trap attraction) and number of traps under both calculations, and they are not quantitatively very far apart. It is also clear that TGO is insensitive to the number of insects in terms of average expectation- this is in agreement observations from several researchers who have used the model, and what one would expect from averaging. Variances may be affected, however. 

\begin{figure}[h]
\begin{center}
\includegraphics[width=4.5in]{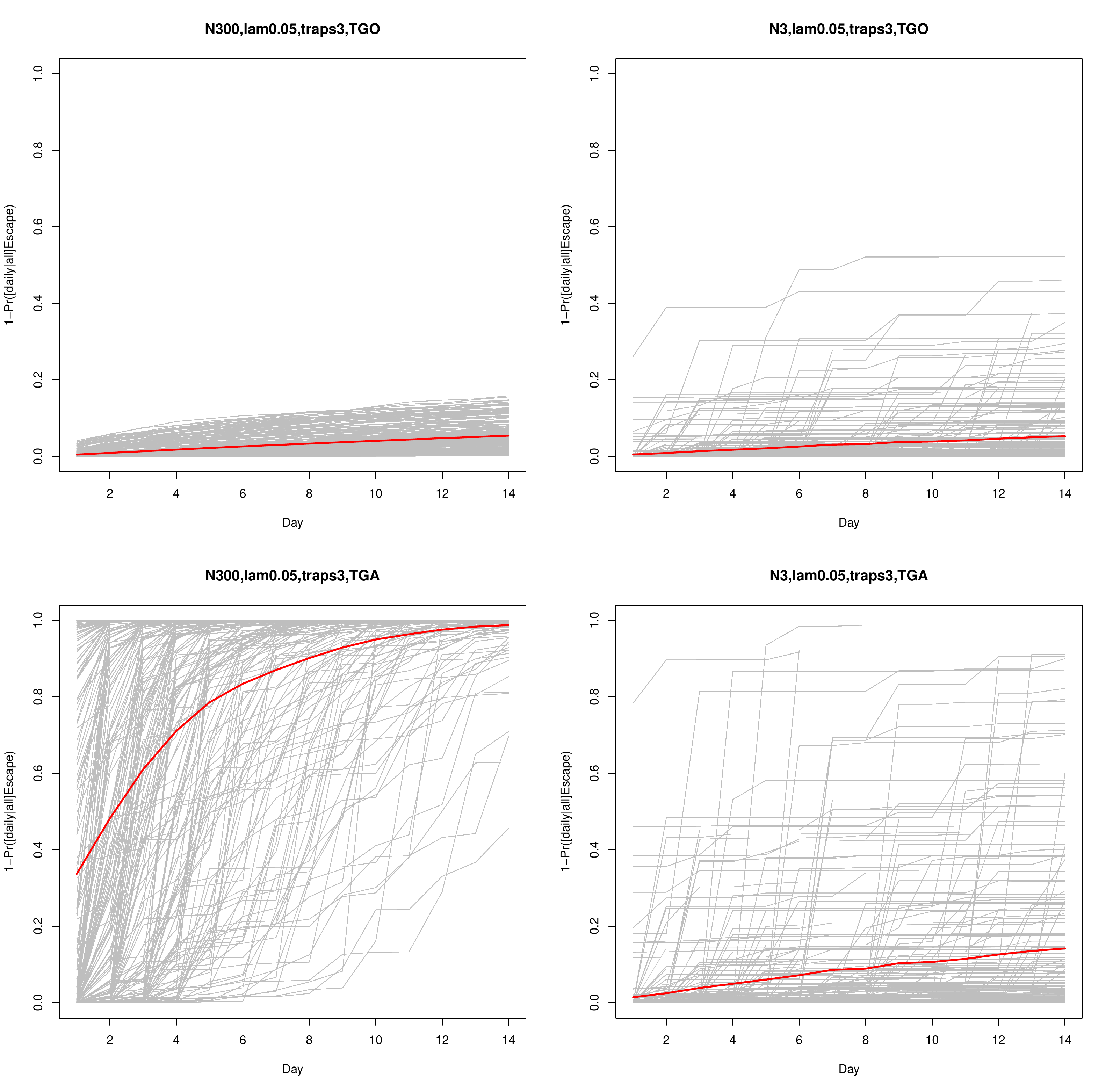}
\caption{Relationship between day and capture, individual runs (grey lines) and mean (red lines) for two algorithms (TGO and TGA) and two numbers of insects ($N$ = 3 or 300). Shared parameters: $\lambda=0.05$, Number of traps = 3}
\label{fig:closeup}
\end{center}
\end{figure}

Fig. \ref{fig:closeup} shows individual run results and means to see how the two algorithms behave relative to numbers of insects. TGO behaves as we have seen: fewer insects increases variation between runs, but the means are unaffected. For TGA there is overall more variation compared with TGO, and it is especially pronounced when there are 3 insects compared with 300; means are strongly affected by the number of insects.

Finally, Fig. \ref{fig:contour} gives a comprehensive comparison how capture probabilities under each algorithm are affected by varying trap attraction and number of traps in the grid. The parameters used follow Fig. 6 in \cite{Manoukis2014Computer}. This comparison shows that for the number of individuals used in the simulation ($N=300$), TGA rapidly saturates to a capture probability of $\sim$ 1. 

\section{Discussion}

The original TrapGrid model calculates the \emph{cumulative daily average escape probability} over time. This is a useful measure for estimating and comparing the sensitivity of trapping networks, especially for cases of surveillance trapping. One feature of this approach is that average values are insensitive to the number of insects in the simulated population, though variance can be affected. In fact, ``insects'' in this mode are better considered as ``sampling points'' rather than literal individuals. TGO is indicated for general comparisons of trap network sensitivity. 

\begin{figure}[h]
\begin{center}
\includegraphics[width=6in]{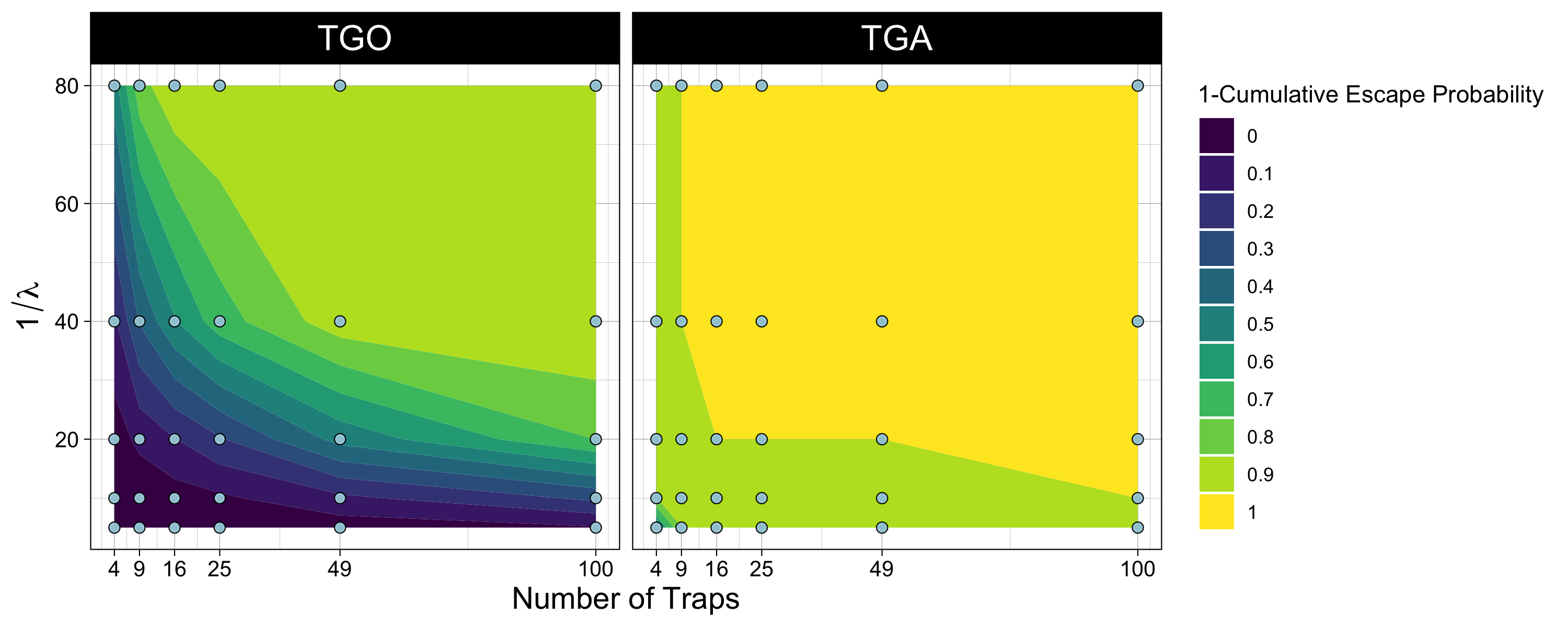}
\caption{Comparison of TGO and TGA using similar parameters as Figure 6 from \cite{Manoukis2014Computer}.  Trap grids were set out on a square mile at distances of 900, 600, 500, 400, 250, and 175 metres,  corresponding to totals of 4,  9,  16,  25,  49 and 100 traps, respectively.  For each trap number, scenarios were run using $1/\lambda$ values of 80, 40, 20, 10,  and 5 metres.  Each simulation had 300 insects randomly placed in a single outbreak location, a $D$ value of $10^4$ and we ran each combination of parameters 250 times.}
\label{fig:contour}
\end{center}
\end{figure}

In instances where population sizes are small and known, the TGA calculation may be useful. TGA is a stricter simulation of detection: it is the \emph{cumulative daily probability of one or more captures}. However, it is likely to saturate quickly for simulated populations over 5-10 individuals, and so may not be as helpful for comparing the relative sensitivity of different trap networks. In other words, the probability of capturing one or more individual generally grows quickly, reaching $\approx$ 1, and then becomes uninformative (see Fig. \ref{fig:contour}).  

As a specific example, assume the size of a population in a small orchard is known with some confidence to be less than 30 individuals. If the user would like to know the probability that none of these insects is captured over the course of two weeks, TGA can provide an estimate of that value for a given set of traps in space and given assumptions about the distribution and movement of individual insects. Such an estimate is not possible with TGO, which would give the average probability of capture. Matching the biology of the insect in question and considerations arising from small arena size (e.g. parametrizing movement via diffusion or random correlated walk) is essential for meaningful application of TrapGrid in small orchards and will require careful thought by the user.

If used for detection network design TGA might be considered unconservative or uninformative. Moreover, deviations in the biology may lead to poor predictions. Though we have not examined the question in detail, it is possible TGO is more robust to variations in \emph{D} or $\lambda$; One reason to expect this is that stochastic effects will play a larger role in the small population scenarios that would be best addressed by TGA. On the other hand, interpretability of TGO in situations where the pest is established or where population sizes are small may not be vary helpful for management decisions. 

In conclusion, we believe that the addition of the TGA calculation to TrapGrid makes the package more versatile and useful for a larger variety of trapping scenarios. It is important to understand in detail the differences between the two calculations for proper interpretation of the results. As mentioned above, a version of TrapGrid (compiled program and source code) with the two modes included is available at the following URL: \url{https://github.com/manoukis/TrapGrid}. 

\section{Acknowledgements}

Thanks to Dr. Peter Follett (USDA-ARS) for comments on an early version of the draft, and to Dr. Peter Caley (CSIRO) for productive discussions. USDA is an equal opportunity provider and employer.

\begin{footnotesize}
\bibliographystyle{unsrtnat}
\bibliography{refs}  
\end{footnotesize}






\section{Appendix}
Java code modified to switch from TGO to TGA in the file Simulation.java. 
\begin{footnotesize}
\begin{lstlisting}
/**
 * Main method for simulation; also writes results to output files.
 */
public SimulationResultsHolder runSimulation() {
	resultsHolder = new SimulationResultsHolder();
	resultsHolder.addFlyReleaseInfo(fr.toString());
	for (int i = 1; i<=numberOfDays; i++) {
		System.err.println("Running simulation for day " + i + "...");
		//double totalProbForDay = 0; //TGO
		double totalProbForDay = 1;  //TGA
		//int numberOfFlies = 0;  //TGO
		Iterator<OutbreakLocation> releasePointItr = fr.allOutbreakLocations.iterator();
		while (releasePointItr.hasNext()) {
			OutbreakLocation currentReleasePoint = releasePointItr.next();
			ArrayList<Point2D.Double> flyLocations = currentReleasePoint.locateFlies(i);
			Iterator<Point2D.Double> flyLocationItr = flyLocations.iterator();
			while (flyLocationItr.hasNext()) {
				Point2D.Double currentLocation = flyLocationItr.next();
				Double currentEscapeProb = tg.getTotalEscapeProbability(currentLocation);
				String[] results = {Integer.toString(i), currentReleasePoint.shortString(),
						locationToString(currentLocation), Double.toString(currentEscapeProb)};
				resultsHolder.addRawData(results);
				//totalProbForDay += currentEscapeProb; //TGO
				totalProbForDay *= currentEscapeProb; //TGA
				//numberOfFlies += 1;  //TGO
			}				
		}				
		//double avgForDay = totalProbForDay / numberOfFlies; //TGO
		double avgForDay = totalProbForDay; //TGA
		cumulativeProb *= avgForDay;
		resultsHolder.addAvgEscapeProbability(i, avgForDay);
	}					
	System.err.println("Simulation complete!");
	return resultsHolder;
}
\end{lstlisting}
\end{footnotesize}

\end{document}